\documentstyle[preprint,aps,epsfig]{revtex}
\draft
\begin{document}
\title{Uranium on uranium collisions at relativistic energies}
\bigskip
\author{Bao-An Li\footnote{Email: Bali@navajo.astate.edu}}
\address{Department of Chemistry and Physics\\
Arkansas State University, P.O. Box 419\\
Jonesboro, Arkansas 72467-0419, USA}

\maketitle

\begin{quote}
Deformation and orientation effects on compression, elliptic 
flow and particle production in uranium on uranium collisions (UU) 
at relativistic energies are studied within the transport 
model ART. The density compression in tip-tip UU collisions is 
found to be about 30\% higher and lasts approximately 50\% longer 
than in body-body or spherical UU reactions. The body-body 
UU collisions have the unique feature that the nucleon elliptic 
flow is the highest in the most central collisions
and remain a constant throughout the reaction. We point out that
the tip-tip UU collisions are more probable to create the QGP at AGS
and SPS energies while the body-body UU collisions are more 
useful for studying properties of the QGP at higher energies.       
\end{quote}
\noindent{PACS number(s):25.75.+r}
\newpage

To better understand the $J/\psi$ suppression mechanism in ultra-relativistic
heavy-ion collisions, uranium on uranium (UU) collisions has been 
proposed recently to extend beyond Pb+Pb collisions at the CERN's 
SPS\cite{henning,lou,satz,ed99a}. Many other outstanding 
issues regarding the corrections to hard processes, the relation 
between elliptic flow and equation of state, as well as 
the study of QCD tri-critical point may also be resolved 
by studying deformation and orientation effects in UU collisions at 
relativistic energies\cite{ed99a,ed99b}. One of the most critical 
factors to all of these issues is the maximum achievable energy 
density in UU collisions. Because of the deformation, UU 
collisions at the same beam energy and impact parameter but 
different orientations are expected to form dense matter 
with different compressions and lifetimes. In particular, the 
deformation of uranium nuclei lets one gain particle 
multiplicity and energy density by aligning the two nuclei
with their long axes head-on (tip-tip). Based on a schematic 
mass scaling of the energy density in relativistic heavy-ion 
collisions\cite{mat}, Braun-Munzinger found a factor of 1.8 
gain in energy density in the tip-tip UU collisions compared to 
the central Au+Au reactions\cite{peter}. More recently, Shuryak 
re-estimated this factor and found it is about 1.3 using 
particle production systematics and geometrical 
considerations of relativistic heavy-ion collisions\cite{ed99a,ed99b}. 
Using a simple Monte-Carlo model, Shuryak has also 
demonstrated that the orientation and the impact parameter 
between the two colliding uranium nuclei can be 
determined simultaneously using the experimentally accessible 
criteria\cite{ed99a,ed99b}. Given the exciting new physics 
opportunities with UU collisions and the obvious discrepancy 
in the estimated gain of energy density , more quantitative studies
with more realistic models are necessary. In this Rapid Communication, 
we report results of such a study. Besides a critical examination 
of the achievable density compression, we also study the nucleon 
elliptic flow and particle production in UU collisions with 
different orientations. 

Our study is based on the relativistic transport model ART for
heavy ion collisions. We refer the reader to Ref. \cite{art} for 
details of the model and its applications in studying various 
aspects of relativistic heavy-ion collisions at beam energies 
from 1 to 20 GeV/A. Uranium is approximately an ellipsoid with 
a long and short semi-axis
\begin{equation}
R_l=R\cdot(1+\frac{2}{3}\delta) 
\end{equation}
and
\begin{equation}
R_s=R\cdot(1-\frac{1}{3}\delta), 
\end{equation} 
where $R$ is the equivalent spherical radius and $\delta$ 
is the deformation parameter. For $^{238}U$, 
one has $\delta=0.27$ and thus a long/short 
axis ratio of about 1.3\cite{bohr}. 

We have performed a systematic study of UU collisions at beam 
energies from 1 to 20 GeV/nucleon. We found similar 
deformation/orientation effects in the whole energy range studied. 
Typical results at beam energies of 10 and 20 GeV/nucleon will 
be presented in the following. Among all possible orientations 
between two colliding uranium nuclei, the tip-tip 
(with long axes head-on) and body-body (with short axes 
head-on and long axes parallel) collisions are the most 
interesting ones\cite{henning,ed99b,peter}. 
Shown in Fig.~1 are the evolution of central baryon densities 
in the UU collisions at a beam energy of 20 GeV/nucleon 
and an impact parameter of $0$ and $6$ fm, respectively. 
In these calculations the cascade mode of the ART model is
used. For comparisons we have also included results for collisions 
between two gold or spherical uranium nuclei.
Indeed, it is interesting to notice that the tip-tip 
UU collisions not only lead to higher compressions but also longer
reaction times. While the body-body UU collisions lead to density 
compressions comparable to those reached in the Au-Au and spherical 
UU collisions. More quantitatively, a 30\% more compression
is obtained in the tip-tip UU collisions at both impact parameters. 
The high density phase (i.e., with $\rho/\rho_0\geq 5$) in the 
tip-tip collisions lasts about 3-5 fm/c longer than the body-body 
collisions. We have seen the same deformation and orientation effects
in the total energy density which also include the newly produced 
particles. The higher compression and longer passage time render 
the tip-tip UU collisions the most probable candidates to form 
the Quark-Gluon-Plasma (QGP) at beam energies that are not very high, 
such as those currently available at the AGS/BNL and SPS/CERN.
  
At RHIC/BNL and LHC/CERN energies, the energy densities in 
colliding spherical heavy nuclei (e.g., Au and Pb) are already
far above the predicted QCD phase transition density. A 30\% 
increase in energy density due to deformation is probably not as 
critical as in fixed-target experiments at lower beam energies. 
A more important issue is how to detect signatures of the 
QGP and extract its properties. How the deformation of uranium nuclei
may help to address this issue? To answer this question we have 
studied the nucleon elliptic flow in UU collisions with different 
orientations. Although our studies are only performed in the beam
energy range of 1-20 GeV/nucleon, the deformation and orientation 
effects are found to be rather energy independent. Our results 
thus may have udeful implications to heavy-ion collisions at even 
higher energies. The elliptic flow reflects the anisotropy in the 
particle transverse momentum ($p_t$) distribution at midrapidity, 
i.e., 
\begin{equation}
v_2= <(p_x^2-p_y^2)/p_t^2>,
\end{equation} 
where $p_x~(p_y)$ is the transverse momentum in (perpendicular to ) 
the reaction plane and the average is taken over all particles in 
all events \cite{oll92,oll98,vol98}. The $v_2$ results from a 
competition between the ``squeeze-out'' perpendicular to the 
reaction plane and the ``in-plane flow''. It has been shown recently 
in many studies that the elliptic flow is particularly sensitive to 
the equation of state of dense 
matter\cite{hung95,sorge97,pawel98,likflow,binel,hei99,liu99,sne99,ed99c}.
Thus, the analysis of $v_2$ is one of the most promising tools 
for detecting signatures of the QGP and extracting its 
properties\cite{e877,liu98,na49,pin99,star,vol99}.

Shown in Fig.~2 are the evolution of the nucleon elliptic flow in
UU collisions with different orientations at a beam energy of 
10 GeV/nucleon and an impact parameter of 6 fm. We initialized
the two uranium nuclei such that their long axes are in the reaction
plane in both tip-tip and body-body collisions. It is seen that 
both the tip-tip and sphere-sphere collisions 
lead to a strong ``in-plane flow'' (positive $v_2$) while 
the body-body reactions result in a large ``squeeze-out'' 
(negative $v_2$). The tip-tip and sphere-sphere collisions 
can't sustain the higher $v_2$ created around the maximum 
compression. This is due to the strong subsequent competition 
between the ``in-plane flow'' and ``squeeze-out'' of baryons. 
While for the body-body collisions the ``squeeze-out'' phenomenon 
dominates throughout the whole reaction because of the strong 
shadowing of matter in the reaction plane. The $v_2$ in body-body 
UU collisions can therefore sustain its early value. The elliptic 
flow in body-body UU collisions is therefore a better probe of 
the high density phase. This point is seen more clearly in the 
impact parameter dependence of the elliptic flow as shown in Fig.~3. 
Unique to the body-body UU collisions, the strength of elliptic 
flow is the highest in the most central collisions where the 
shadowing effect in the reaction plane is the strongest. 
While in tip-tip and sphere-sphere UU collisions the elliptic 
flow vanishes in the most central collisions due to symmetry. 
Therefore, the ``squeeze-out'' of particles including newly 
created ones perpendicular to the reaction plane in very central 
body-body UU collisions can provide direct information about the 
dense matter formed in the reaction. This is clearly an advantage 
of using the body-body collisions over the tip-tip collisions. Of 
course, at collider energies it is more important to study the 
elliptical flow at midrapidities for newly produced particles, 
such as pions which are even more sensitive to the nuclear 
shadowing effects\cite{li93,e895,likflow}. 
 
It is also of considerable interest to study deformation 
and orientation effects on particle production. 
Shown in Fig.~4 are the 
multiplicities of pions and positive kaons as a function of impact 
parameter. The maximum impact parameter for the tip-tip and 
body-body UU collisions are $2R_s$ and $2R_l$, respectively. 
As one expects the central 
(with $b\leq 5$ fm) tip-tip UU collisions produce more particles 
due to the higher compression and the longer passage time of 
the reaction. While at larger impact parameters, the
smaller overlap volume in the tip-tip collisions leads to less 
particle production than the body-body and sphere-sphere reactions. 
Also as one expects from the reaction geometry, the multiplicities 
in the body-body collisions approach those in the sphere-sphere 
collisions as the impact parameter reaches zero. In the most 
central collisions, the tip-tip UU collisions produce about 
15\% (40\%) more pions (positive kaons) than the body-body and 
sphere-sphere UU collisions. These deformation and orientation 
effects on particle production are consistent with those on density 
compression shown in Fig.~1. Compared to pions, kaons are more 
sensitive to the density compression since most of them are 
produced from second chance particle (resonance)-particle 
(resonance) scatterings at the energies studied here\cite{ko}.      
  
In summary, using A Relativistic Transport model we have studied 
the deformation and orientation effects on the compression, elliptic 
flow and particle production in uranium on uranium (UU) collisions 
at relativistic energies. The compression in the tip-tip UU 
collisions is about 30\% higher and lasts approximately 
50\% longer than in the body-body or spherical UU collisions.
Moreover, we found that the nucleon elliptic flow in the 
body-body UU collisions have some unique features. We 
have pointed out that the tip-tip UU collisions are more probable 
to create the QGP at the AGS/BNL and SPS/CERN energies. 
While at RHIC/BNL and LHC/CERN energies, the ``squeeze-out'' 
of particles in the central body-body collisions is more 
useful for studying properties of the QGP.    
  
I would like to thank W.F. Henning for suggesting me to work on this
project and stimulating discussions. I am also grateful to C.M. Ko,
M. Murray, J.B. Natowitz, E.V. Shuryak, A.T. Sustich, and B. Zhang 
for helpful discussions. This work was supported in part by 
a subcontract S900075 from Texas A\&M Research Foundation's NSF 
Grant PHY-9870038.

\newpage

\begin{figure}[h]
\vspace{1in}
\centerline{\epsfig{file=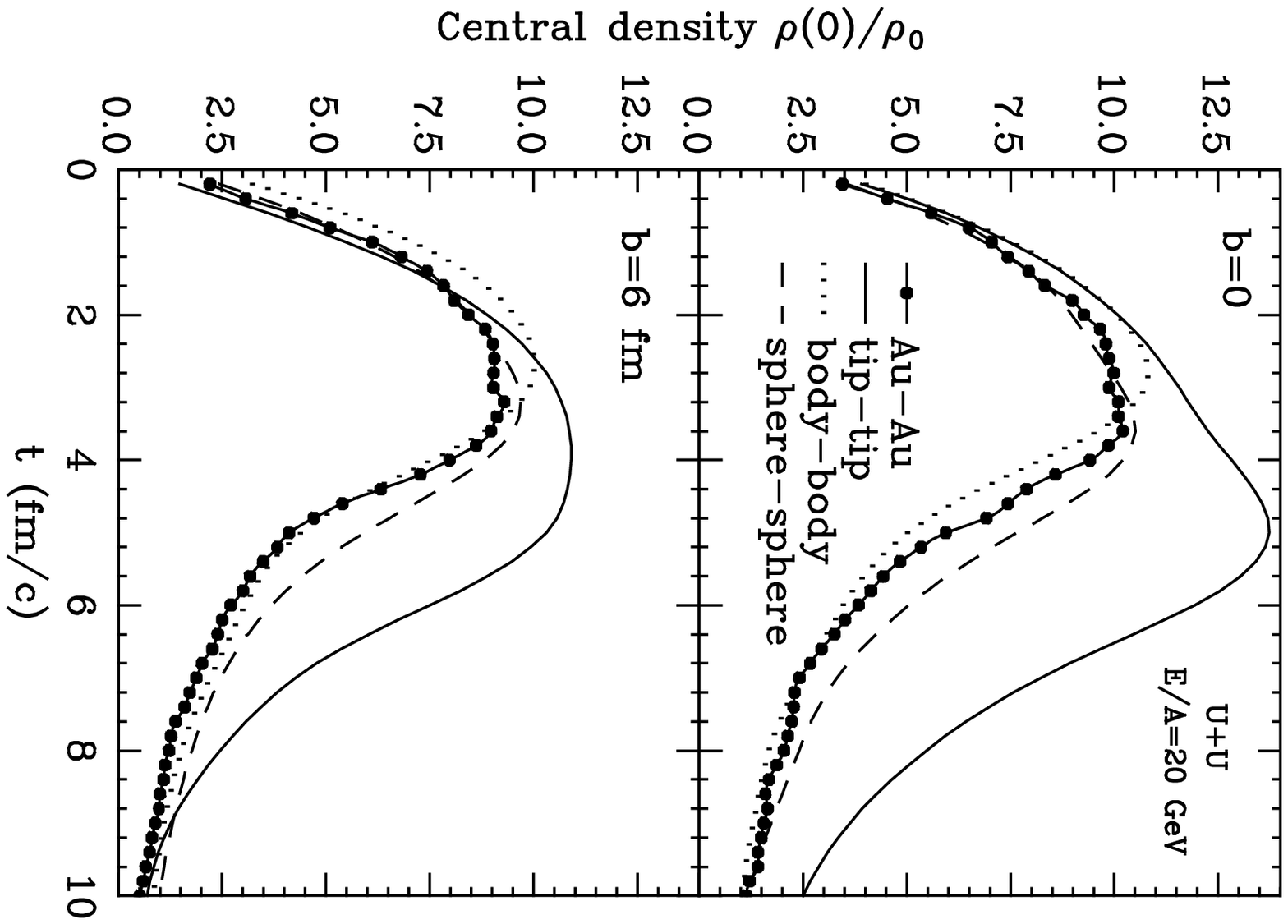,width=4in,height=4in,angle=90}}
\vspace{0.3in}
\caption{The evolution of central baryon density in Au-Au 
(filled circles), tip-tip (solid line), body-body (dotted line) and 
sphere-sphere (dashed line) UU collisions at a beam energy of 20 GeV/nucleon and
an impact parameter of 0 (upper panel) and 6 fm (lower panel), 
respectively.}
\label{fig1} 
\end{figure}

\newpage

\vspace{12in}
\begin{figure}[h]
\centerline{}
\vspace{1in}
\centerline{\epsfig{file=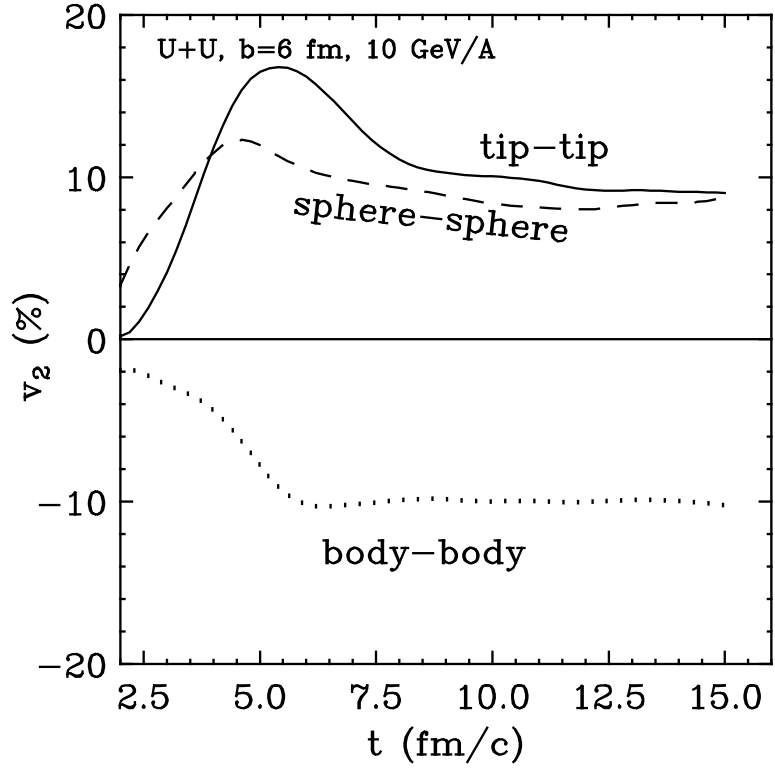,width=4in,height=4in,angle=90}}
\vspace{0.3in}
\caption{The nucleon evolution of elliptic flow in the UU collisions 
at a beam energy of 10 GeV/nucleon and an impact parameter of 6 fm.}
\label{fig2} 
\end{figure}

\newpage

\vspace{12in}
\begin{figure}[h]
\centerline{}
\vspace{1in}
\centerline{\epsfig{file=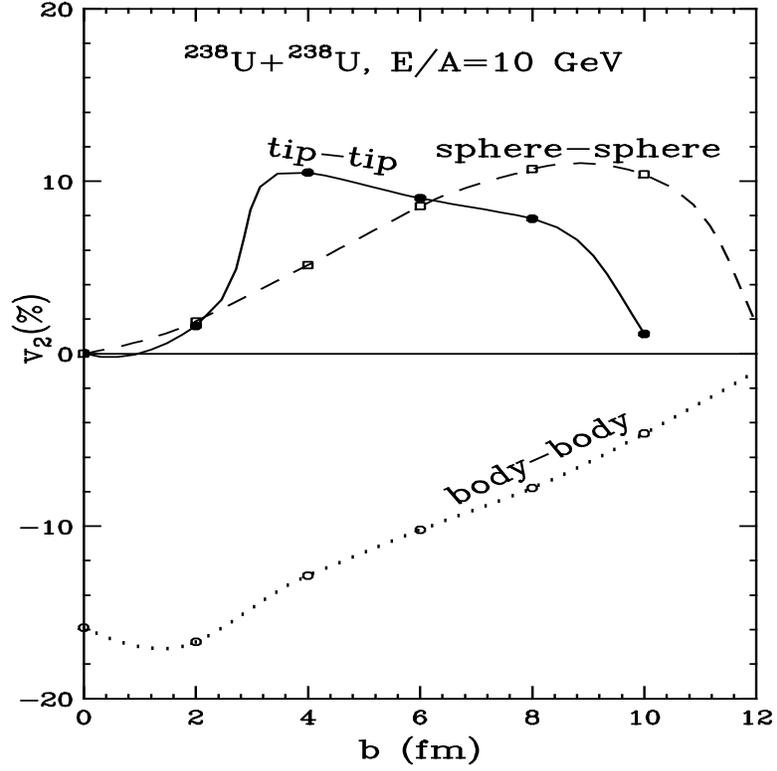,width=4in,height=4in,angle=90}}
\vspace{0.3in}
\caption{The impact parameter dependence of nucleon elliptic flow in the 
tip-tip (solid line), body-body (dotted line) and sphere-sphere (dashed line) 
UU collisions at a beam energy of 10 GeV/nucleon.}
\label{fig3} 
\end{figure}

\newpage

\vspace{12in}
\begin{figure}[h]
\centerline{}
\vspace{1in}
\centerline{\epsfig{file=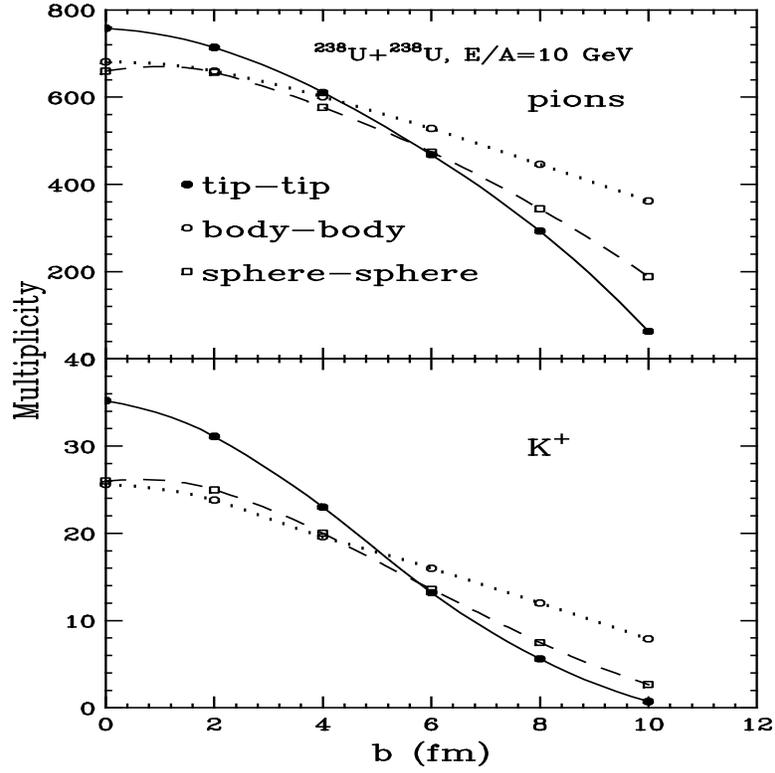,width=4in,height=4in,angle=90}}
\vspace{0.3in}
\caption{The impact parameter dependence of pion (upper panel) 
and positive kaon (lower panel) production in the 
tip-tip (solid line), body-body (dotted line) and sphere-sphere (dashed line) 
UU collisions at a beam energy of 10 GeV/nucleon.}
\label{fig4} 
\end{figure}

\end{document}